\documentclass[aps,graphicx,10pt,showkeys,showpacs,onecolumn]{revtex4}
\usepackage{amsmath}
\usepackage{amscd}
\usepackage{graphicx}
\usepackage{subfigure}
\usepackage{appendix}
\usepackage{multirow}
\usepackage{mathrsfs}

\begin{document}

\title[Short Title]{Enhanced quantum sensing in time-modulated non-Hermitian systems}

\author{Qi-Cheng Wu$^{1,}$\footnote{E-mail: wuqi.cheng@163.com}}
\author{Yan-Hui Zhou$^{1}$}
\author{Tong Liu$^{1}$}
\author{Yi-Hao Kang$^{2}$}
\author{Qi-Ping Su$^{2}$}
\author{Chui-Ping Yang$^{2,}$\footnote{E-mail: yangcp@hznu.edu.cn}}

\affiliation{$^{1}$Quantum Information Research Center and Jiangxi Province Key Laboratory of Applied Optical Technology, Shangrao Normal University, Shangrao 334001, China\\
$^{2}$School of Physics, Hangzhou Normal University, Hangzhou
311121, China}

\begin{abstract}

Enhancing the sensitivity of quantum sensing near an exceptional
point represents a significant phenomenon in non-Hermitian (NH)
systems. However, the application of this property in
time-modulated NH systems remains largely unexplored. In this
work, we propose two theoretical schemes to achieve enhanced
quantum sensing in time-modulated NH systems by leveraging the
coalescence of eigenvalues and eigenstates. We conduct a
comprehensive analysis of the full energy spectrum, including both
real and imaginary components, the population distribution of
eigenstates, and various characteristics associated with optimal
conditions for sensitivity enhancement. Numerical simulations
confirm that eigenvalue-based quantum sensors exhibit a 9.21-fold
improvement compared to conventional Hermitian sensors, aligning
with the performance of existing time-independent NH sensors.
Conversely, for eigenstate-based quantum sensors, the enhancement
reaches up to 50 times that of conventional Hermitian sensors,
surpassing the results of existing time-independent NH sensors.
Moreover, the eigenstate-based sensor exhibits divergent
susceptibility even when not close to an exceptional point. Our
findings pave the way for advanced sensing in time-sensitive
contexts, thereby complementing existing efforts aimed at
harnessing the unique properties of open systems.
\end{abstract}


\maketitle

\section{Introduction}

 Since the discovery of exceptional point
(EP)~\cite{EPs1,EPs2,EPs3}, a wealth of counterintuitive phenomena
and applications have emerged in non-Hermitian (NH)
systems~\cite{NH1,PT,NH2,NH3,topology1,topology-encircling}. In
particular, EP-based high-precision sensors play an increasingly
crucial role in modern
science~\cite{sensing-microcavity,Enhanced-sensitivity,quantum-sensing1,
quantum-sensing2,quantum-sensing3,quantum-sensing4,quantum-sensing5,quantum-sensing6,sensing-add1,sensing-add2,sensing-add3,sensing-add4}.
The unique feature of EP-based sensors is the divergent eigenvalue
susceptibility arising from the square-root frequency topology
around exceptional points~\cite{square-root1,square-root2}, which
has been demonstrated in classical optical or electromechanical
systems~\cite{Microcavity-sensor,acoustic-sensor,Sagnac-effect},
trapped ions~\cite{trapped-ions}, solid spins~\cite{single-spin},
single photons~\cite{single-photon}, or qubit-resonator
systems~\cite{qubit-resonator}. Recently, several studies have
indicated that eigenstate coalescence can also lead to divergent
susceptibility in the normalized state population of NH systems,
without relying on the existence of EP.  For instance, Chu
\emph{et al.} proposed a strategy to realize a single-qubit
pseudo-Hermitian sensor, which exhibits divergent susceptibility
in a dynamical evolution not necessarily involving an
EP~\cite{pseudo-hermitian1}. Furthermore, Xiao \emph{et al.}
introduced and experimentally validated a  NH sensing scheme that
enhances sensitivity without depending on
EP~\cite{pseudo-hermitian2}. These schemes offer promising
pathways for realizing efficient quantum sensors in NH domains.

Nevertheless, a common assumption in these
schemes~\cite{sensing-microcavity,Enhanced-sensitivity,quantum-sensing1,
quantum-sensing2,quantum-sensing3,quantum-sensing4,quantum-sensing5,quantum-sensing6,
square-root1,square-root2,Microcavity-sensor,acoustic-sensor,Sagnac-effect,trapped-ions,
single-spin,single-photon,qubit-resonator,pseudo-hermitian1,pseudo-hermitian2,sensing-add1,sensing-add2,sensing-add3,sensing-add4}
is that the Hamiltonian of the designed sensor is
time-independent. In practice, for a broader range of cases, the
Hamiltonian of NH systems is
time-dependent~\cite{time-dependent1,time-dependent2,time-dependent3,time-dependent4,Arkhipov-encircling2},
leading to more complex system dynamics. Furthermore, unlike
sensors governed by time-independent Hamiltonians, deriving
analytical solutions for the dynamic evolution of systems with
time-dependent Hamiltonians presents substantial challenges.
Consequently, pinpointing parameter intervals that yield optimal
sensitivity for estimation purposes becomes considerably more
challenging. On the other hand, intuitively, the incorporation of
time as an additional degree of freedom markedly enhances the
maneuverability of time-dependent NH sensors in comparison to
traditional sensors, which may give rise to distinctive effects.
Therefore, exploring quantum sensing based on time-dependent NH
systems and its potential advantages is highly valuable.

In this work, we introduce an additional temporal dimension to
modulate the sensor's sensitivity to parameter perturbations,
thereby bridging the gap between time-independent and
time-dependent systems. We propose two theoretical schemes that
leverage the coalescence of eigenvalues and eigenstates to achieve
enhanced quantum sensing in time-modulated NH systems. A thorough
analysis of various features associated with optimal conditions
for sensitivity enhancement is conducted, including the optimal
timing for achieving high sensitivity using the energy spectrum of
eigenvalues and the population distribution of eigenstates. When
the time-evolution trajectory of the system approaches the EP,
eigenvalue-based quantum sensors exhibit a 9.21-fold improvement
over conventional Hermitian sensors, aligning with the performance
of existing time-independent NH sensors. In contrast, for
eigenstate-based quantum sensors, the enhancement reaches up to 50
times that of conventional Hermitian sensors, significantly
surpassing the results of existing time-independent NH sensors.
Notably, the eigenstate-based sensor continues to display
divergent susceptibility even when the time-evolution trajectory
of the system deviates from the EP. Our findings open a new avenue
for enhanced sensing in time-sensitive frameworks, complementing
existing efforts to leverage the unique properties of open
systems.

The rest of this paper are organized as follows. In
Sec.~\ref{section:II}, we provide a concise overview of the model
for the time-modulated NH sensor. Sections~\ref{section:III}
and~\ref{section:IV} are dedicated to detailed analyses of the
energy spectrum, eigenstate population distribution, and various
characteristics associated with optimal conditions for sensitivity
enhancement in eigenvalue-based and eigenstate-based quantum
sensors, respectively. In Sec.~\ref{section:V}, we discuss the
feasibility of the experimental realization. Finally, we present a
summary in Section~\ref{section:VI}.

\section{Model of the time-modulated non-Hermitian sensor}\label{section:II}

We consider a generic time-modulated NH qubit sensor subjected to
a weak external field. The overall Hamiltonian is expressed as $H
= H_S(t) + \lambda H'$, where $\lambda$ represents the small
parameter to be estimated, and $H'$ depends on the form of the
weak external field. In an orthonormal basis of states
$\{|1\rangle, |0\rangle\}$, the Hamiltonian of the unperturbed
time-modulated NH sensor can be expressed as (assuming $\hbar=1$)
\begin{eqnarray}\label{eq-Hs}
H_{S}(t)=\left(\begin{array}{ccccccc}
\epsilon+\delta(t)-\frac{i\gamma_{-}(t)}{2} &  g\\
g & \epsilon-\frac{i\gamma_{+}(t)}{2} \\
\end{array}\right),
\end{eqnarray}
where $\epsilon+\delta(t)$ and $\epsilon$ are the energies of the
two levels $|1\rangle$, $|0\rangle$, and $g$ is the interaction
strength between the two levels.  $\gamma_{+}(t)$ and
$\gamma_{-}(t)$ are ``gain'' and ``loss'' rates (dissipative
rates), respectively. When $\gamma_{+}(t)=\gamma_{-}(t)=0$, the NH
Hamiltonian $H_{S}(t)$ reduces to a Hermitian Hamiltonian
\begin{eqnarray}\label{eq-H0}
H_{0}(t)=\left(\begin{array}{ccccccc}
\epsilon+\delta(t) &  g\\
g & \epsilon \\
\end{array}\right).
\end{eqnarray}

After performing algebraic calculations, one can find that the
eigenvalues of the NH Hamiltonian $H_{S}(t)$ are
\begin{eqnarray}\label{eq1-3a}
E_{\pm}(t)=\frac{1}{4}[2(2\epsilon+\delta(t))-i(\gamma_{+}(t)+\gamma_{-}(t))\pm{2\Delta_E(t)}],
\end{eqnarray}
with
\begin{eqnarray}\label{eq1-3}
\Delta_E(t)=\frac{i}{2}\sqrt{[\gamma_{-}(t)-\gamma_{+}(t)+2i\delta(t)]^2-16g^{2}}.
\end{eqnarray}
The corresponding right eigenvectors  are
\begin{eqnarray}\label{eq1-4}
|\phi_{+}(t)\rangle&=&\frac{1}{S_{+}(t)}[A_{+}(t),4g]^{T},\cr
|\phi_{-}(t)\rangle&=&\frac{1}{S_{-}(t)}[A_{-}(t),4g]^{T},
\end{eqnarray}
where
$A_{\pm}(t)=2\delta(t)-i[\gamma_{-}(t)-\gamma_{+}(t)]\pm{2\Delta_E}(t)$
and $S_{\pm}(t)=\sqrt{|A_{\pm}(t)|^2+|4g|^2}$,  together with the
left eigenvectors
\begin{eqnarray}\label{eq1-5}
\langle\widehat{\phi_{+}}(t)|&=&\frac{1}{S_{+}(t)}[A_{+}(t),4g],\cr
\langle\widehat{\phi_{-}}(t)|&=&\frac{1}{S_{-}(t)}[A_{-}(t),4g].
\end{eqnarray}
The biorthogonal partners $\{\langle\widehat{{\phi_{n}}}|\}$,
$\{|{\phi}_{m}\rangle \}$ $(n,m$=$+,-)$ are normalized to satisfy
the biorthogonality and closed relations
$\langle\widehat{{\phi_{n}}}(t)|\phi_{m}(t)\rangle$=$\delta_{nm}$,
$\sum_{n}|\widehat{{\phi_{n}}}(t)\rangle\langle\phi_{n}(t)|$=$\sum_{n}|{\phi_{n}}(t)\rangle\langle\widehat{{\phi_{n}}}(t)|$=1~\cite{NH3}.

\section{eigenvalue-based quantum sensing}\label{section:III}

Without loss of generality,  the form of the perturbation
Hamiltonian ${H'}$ can be set as ${H'}=\sigma_{x}$, and the
estimated parameter $\lambda$ can be considered as deviations of
the interaction strength $g$. In this section, we will propose a
theoretical framework for eigenvalue-based quantum sensing to
detect minute deviations $\lambda$ of the interaction strength $g$
from its normal value $g_{0}$, defined as $\lambda=g-g_{0}$.

Based on Eq.~(\ref{eq1-3}), the derivative of the energy splitting
$\Delta_{E}(t)$ to the parameter $g$ at time $t$ (i.e., the
susceptibility of the quantum sensor) is
\begin{eqnarray}\label{eq1-6}
\chi(t)=\frac{\partial\Delta_E(t)}{\partial{g}}=\frac{4g}{\Delta_E(t)},
\end{eqnarray}
which diverges under an eigenenergy-matching condition
[$\Delta_E(t)\rightarrow0$ near EP or diabolic point (DP)]. This
divergent behavior implies that even a small variation of $g$
causes a significant change of the energy splitting around the EP
or DP at time $t$.

It is important to note that in NH systems, the energy splitting
$\Delta_E(t)$ is generally a complex quantity. However, when
$\gamma_{+}(t) = \gamma_{-}(t) = 0$, the Hermitian Hamiltonian
$H_{0}(t)$ causes $\Delta_E(t)$ to become a real quantity. To
uniformly evaluate the performance of the quantum sensor in both
Hermitian and NH regimes, we characterize the
non-Hermiticity-enabled sensitivity enhancement by
\begin{eqnarray}\label{eq1-7}
S_{E}(t) =|\frac{\chi_{NH}(t)}{\chi_{H}(t)}|,
\end{eqnarray}
where $\chi_{H}(t)$ and $\chi_{NH}(t)$ are the susceptibility of
the quantum sensor in Hermitian and NH regimes, respectively.

\subsection{Optimal time window for achieving high sensitivity}

Unlike quantum sensors based on time-independent systems, those
based on time-dependent systems introduce an additional temporal
dimension that enables the regulation of sensitivity to parameter
perturbations. A comprehensive investigation into the optimal time
window for achieving high sensitivity is crucial. For the sake of
discussion, we assume $\gamma_{+}(t)=\alpha\gamma_{-}(t)$ and
normalize all parameters with respect to the energy scale
$\epsilon$ in the following analysis. The analysis of
Eq.~(\ref{eq1-3}) reveals that exceptional points (EPs) occur
under two conditions: (i) when $\delta = g = \gamma_{-} = 0$, and
(ii) when $\delta = 0$, $\alpha=-1$ and $2g = |\gamma_{-}| \neq
0$. In both cases, the energy discriminant $\Delta_E(\delta, g,
\gamma_{-})$ equals zero. However, simultaneously analyzing the
relationship between the time-dependent parameters $[\delta(t), g,
\gamma_{-}(t)]$ and $\Delta_E(t)$ poses a considerable challenge.
Therefore, it is advisable to fix certain parameters and conduct
an in-depth investigation into the correlation between the
remaining parameters and $\Delta_E(t)$.

Theoretically, $\Delta_E(t)$ can be rewritten as
\begin{eqnarray}\label{eq1-8}
\Delta_E(t)&=&\rho(t)\exp(\frac{i[\theta(t)+\pi]}{2}),
\end{eqnarray}
with
\begin{eqnarray}\label{eq1-8}
\rho^2(t)&=&[c(t)-a(t)]^2+2b(t)[c(t)-a(t)]+b^2(t)+4a(t)b(t),\cr
\theta(t)&=&\arctan[\frac{2\sqrt{a(t)b(t)}}{a(t)-b(t)-c(t)}],
\end{eqnarray}
where $a(t)=(1-\alpha)^2\gamma^{2}_{-}(t)$, $b(t)=4\delta^2(t)$,
and $c(t)=16g^2$. Then, the eigenvalue-matching condition
$\Delta_E(t) \rightarrow 0$ can be equivalently transformed into
the condition $\rho^2(t) \rightarrow 0$. From Eq.~(\ref{eq1-8}),
it is evident that $\rho^2(t)$ can be expressed as a
time-dependent parabola with an upward opening. To analyze the
impact of changes in the parameter $g$ on $\rho^2(t)$, we consider
$c(t) - a(t)$ as a variable. Consequently, $\rho^2(t)$ attains its
minimum value $\rho^2(\tau)_{\textrm{min}}$ at time $\tau$, where
$c(\tau) - a(\tau) = -b(\tau)$. We emphasize that
$|\rho(\tau)|_{\textrm{min}} = 0$ implies $\Delta_E(\tau) = 0$,
indicating that the system is at an EP or DP. Consequently,
$|\rho(\tau)|_{\textrm{min}}$ can serve as a quantitative measure
to evaluate the deviation of the system's parameter trajectory
from the EP or DP.

In addition, if $c(t)$ is a positive constant coefficient $c$ and
$0 < c \leq \max[a(t) - b(t)]$, the existence of an optimal time
$\tau$,
\begin{eqnarray}\label{eq1-10}
c=[a(\tau)-b(\tau)],
\end{eqnarray}
can be guaranteed, ensuring high sensitivity of $\rho^2(t)$ to
changes in the parameter $c$. However, if \( c > \max[a(t) - b(t)]
\), the relationship \( c - a(\tau) = -b(\tau) \) cannot hold. In
this scenario, a notably high sensitivity of \( \rho^2(t) \) to
variations in the parameter \( c \) remains evident when
\begin{eqnarray}\label{eq1-11}
|a(\tau) - b(\tau)|= \max[a(t) - b(t)].
\end{eqnarray}
It is important to note that when the parameter $c$ becomes
significantly large, the trajectory of the system parameter change
deviates substantially from the EP, resulting in
$|\rho(\tau)|_{\textrm{min}} \gg 0$. Consequently, the
eigenvalue-matching condition no longer holds, leading to a rapid
decrease in the susceptibility of the quantum sensor as $c$
increases. Therefore, if the forms of $a(t)$ and $b(t)$ are fixed,
for the interaction strength $g$, there exists a trade-off between
achieving high sensitivity [as shown in Eq.~(\ref{eq1-6}), where
$\chi(t)$ is proportional to $g$] and maintaining the
eigenvalue-matching condition over time [$0 < 16g^2 < \max|a(t) -
b(t)|$].

\subsection{Performance of eigenvalue-based quantum sensing}

\begin{figure}[htb]
\scalebox{0.46}{\includegraphics{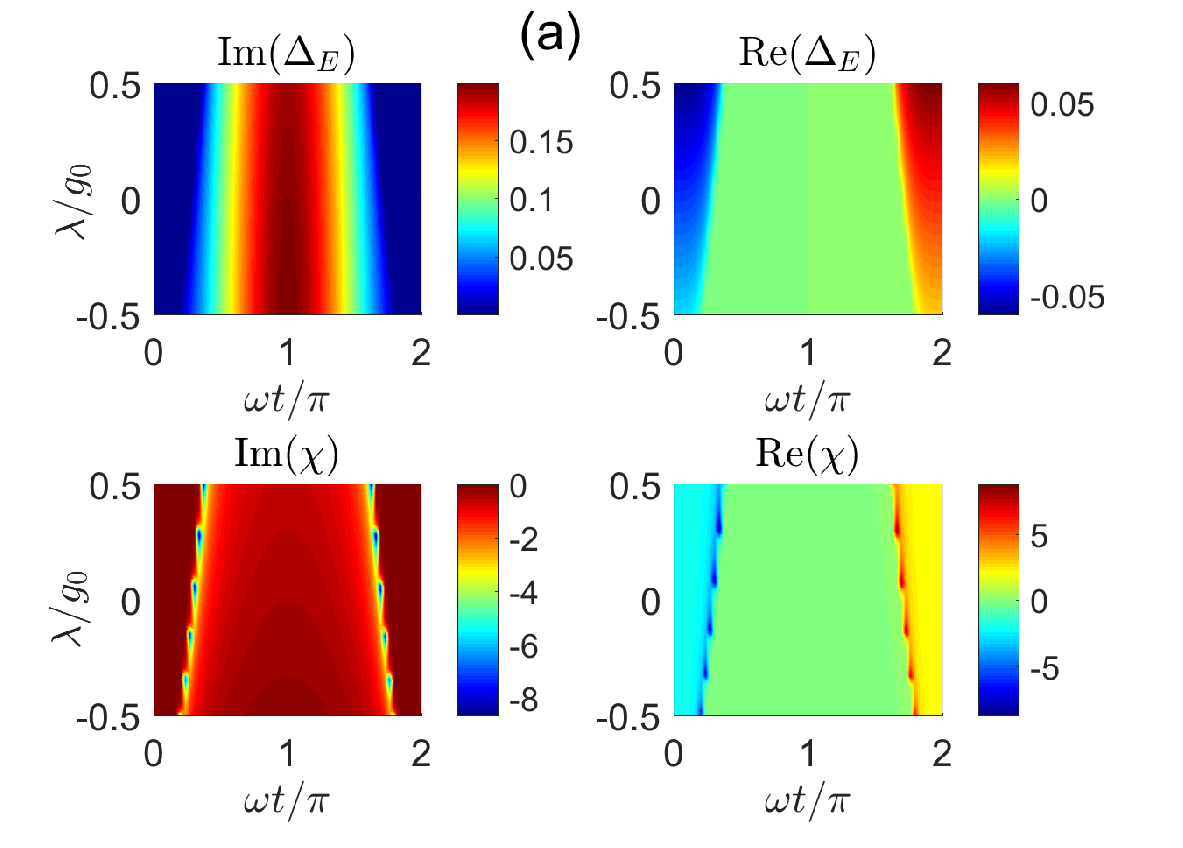}\includegraphics{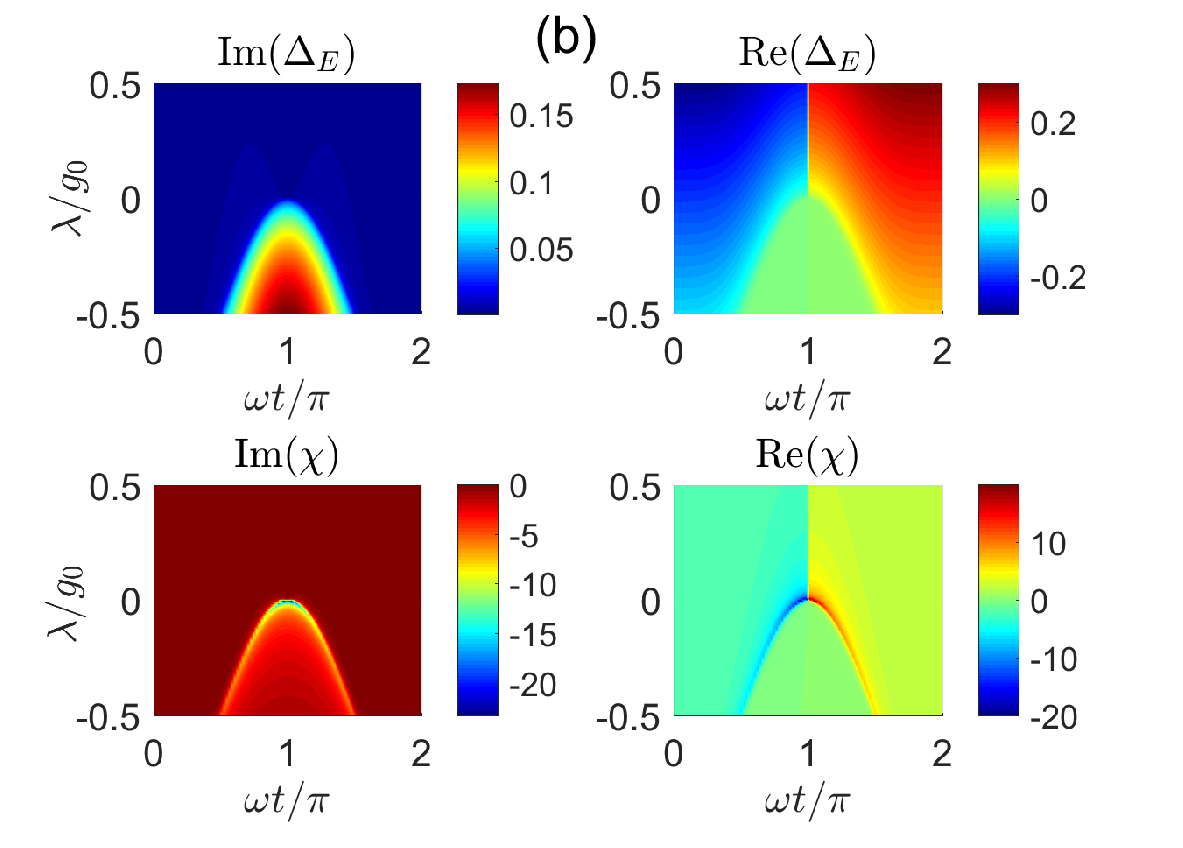}}
\caption{\label{fig-Evstg} The real  and imaginary  parts of the
energy splitting $\Delta_E(t)$ and the susceptibility $\chi(t)$ as
functions of $\lambda/g_{0}$ and $\omega t/\pi$ for the NH sensor.
In panels (a) and (b), the interaction strength $g_{0}$ is set to
0.02 and 0.1, respectively.  Other parameters are fixed at
$\Delta_{0}=0.04$, $\alpha=-1$, and $\Gamma_{0}=0.2$.}
\end{figure}

To gain a more intuitive understanding of the aforementioned
results, let us examine a time-modulated parameter
trajectory~\cite{wu-encircling,bell-encircling}
\begin{eqnarray}\label{eq1-12}
g=g_{0},\delta(t)=\Delta_{0}\sin(\omega{t}),\gamma_{-}(t)=\Gamma_{0}\sin^2(\frac{\omega{t}}{2}),
\end{eqnarray}
where $g_{0}$, $\Delta_{0},\Gamma_{0},$  $\omega \in \mathbf{R}$
are constants, the period of the trajectory is $T=2\pi/\omega$. To
assess the validity of eigenvalue-based quantum sensing, we
investigate the time-dependent behavior of the complex parameters
(i.e., the energy splitting $\Delta_E(t)$ and the susceptibility
$\chi(t)$) by analyzing the influence of the perturbation
$\lambda/g_{0}$.

As illustrated in Fig.~\ref{fig-Evstg}(a), both the real and
imaginary components of the energy splitting $\Delta_E$ exhibit
non-monotonic variations, characterized by abrupt transitions at
specific intervals (from blue to green, and from green to red).
Correspondingly, sensitivity peaks for $\textrm{Im}(\chi)$ are
observed in the deep blue region; while for $\textrm{Re}(\chi)$,
they occur in both the deep blue and red regions. This divergent
behavior indicates that a minor variation in $\lambda/g_{0}$ can
lead to significant changes in the splitting at specific times
$\tau$. Consequently, by selecting appropriate time nodes, we can
achieve optimal sensitivity detection for different disturbances
$\lambda/g_{0}\approx \Delta_g$ (where $\Delta_g$ is an arbitrary
constant coefficient). Alternatively, we can finely adjust the
magnitude of $g_0$ so that the sensitivity peak occurs when
$\lambda/g_{0}\approx 0$ for a fixed time. Similarly, as
illustrated in Fig.~\ref{fig-Evstg}(b), when $g_{0}=0.1$, the
energy splitting $\Delta_E$ undergoes a phase transition at the
point ($\lambda/g_{0}=0$, $t=\pi/\omega$). At this critical point,
sensitivity peaks with $|\textrm{Im}(\chi)| \approx 22$ and
$|\textrm{Re}(\chi)| \approx 20$ are also observed.

\begin{figure}[htb]
\scalebox{0.5}{\includegraphics{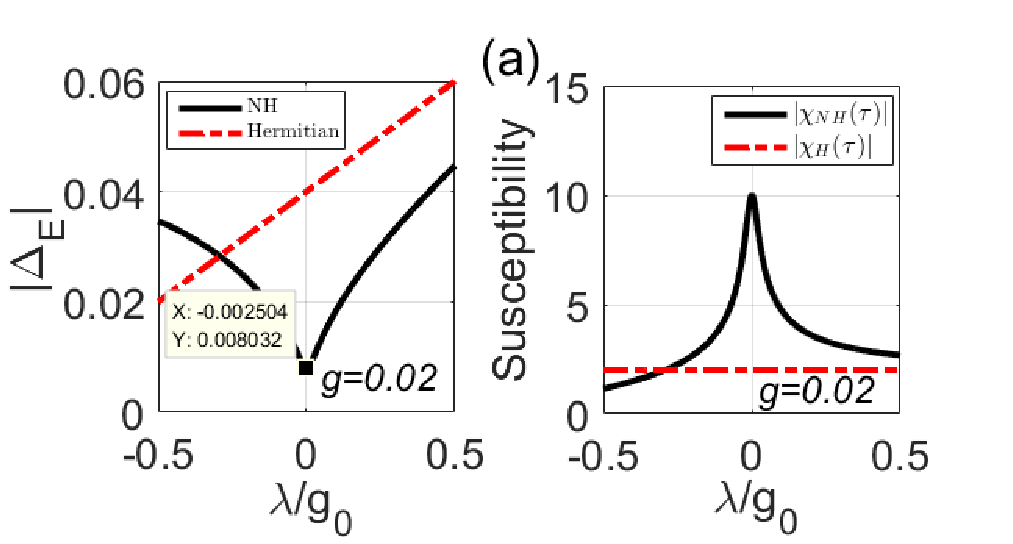}}
\scalebox{0.5}{\includegraphics{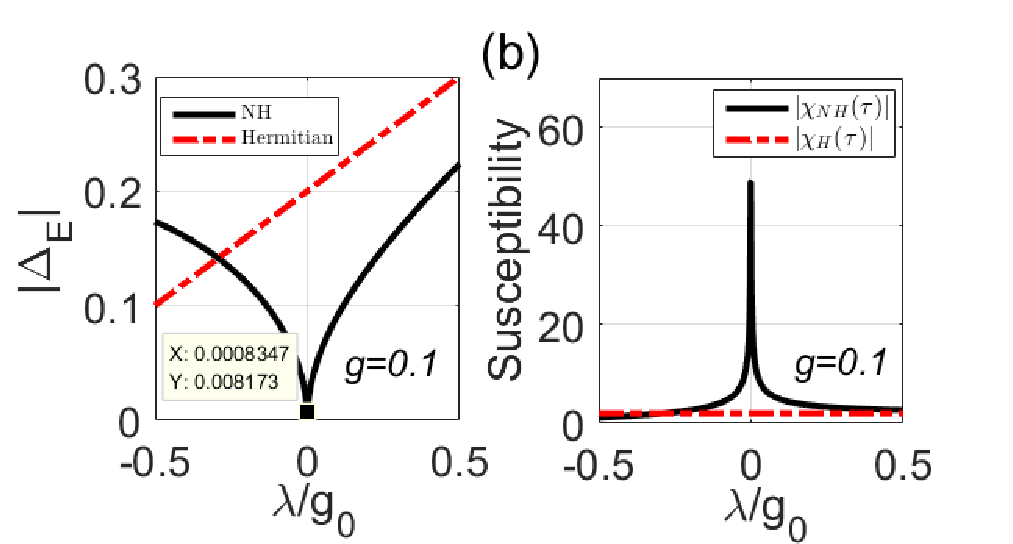}}
\caption{\label{fig-Evsg}  The energy splitting $\Delta_E(\tau)$
and the susceptibility $|\chi_{NH}(\tau)|$ [$|\chi_{H}(\tau)|$] as
functions of $\lambda/g_0$ for the NH sensor (solid black line)
and its Hermitian counterpart (dotted red line), respectively. The
interaction strength $g_0$ and the selected time $\tau$ are chosen
as 0.02 and 0.295$\pi/\omega$ in panel (a), while  0.1 and
$\pi/\omega$ in panel (b) according to Eq.~(\ref{eq1-10}). Other
parameters are set as follows: $\Delta_0 = 0.04$, $\alpha = -1$,
$\Gamma_0 = 0.2$ for the NH sensor, and $\Gamma_0 = 0$ for the
Hermitian sensor.}
\end{figure}

Next, we investigate the energy splitting \( |\Delta_E(t)| \) and
the susceptibility $|\chi_{NH}(\tau)|$ $[|\chi_{H}(\tau)|]$ for
both the NH sensor and its Hermitian counterpart under varying
parameters. For the NH sensor, the time-dependent functions \(
\delta(t) \) and \( \gamma(t) \) are, respectively, defined as \(
\delta(t) = 0.04\sin(\omega t) \) and \( \gamma(t) =
0.2\sin^2\left(\frac{\omega t}{2}\right) \) for simplicity. In
contrast, for the Hermitian sensor, \( \delta(t) = 0.04\sin(\omega
t) \) but \( \gamma(t) = 0 \). Figure~\ref{fig-Evsg} illustrates
the energy splitting \( |\Delta_E(t)| \) and the susceptibility
$|\chi_{NH}(\tau)|$ $[|\chi_{H}(\tau)|]$ as functions of \(
\lambda / g_0 \) for both the NH sensor (solid black line) and its
Hermitian counterpart (dotted red line).

\begin{figure}
\scalebox{0.5}{\includegraphics{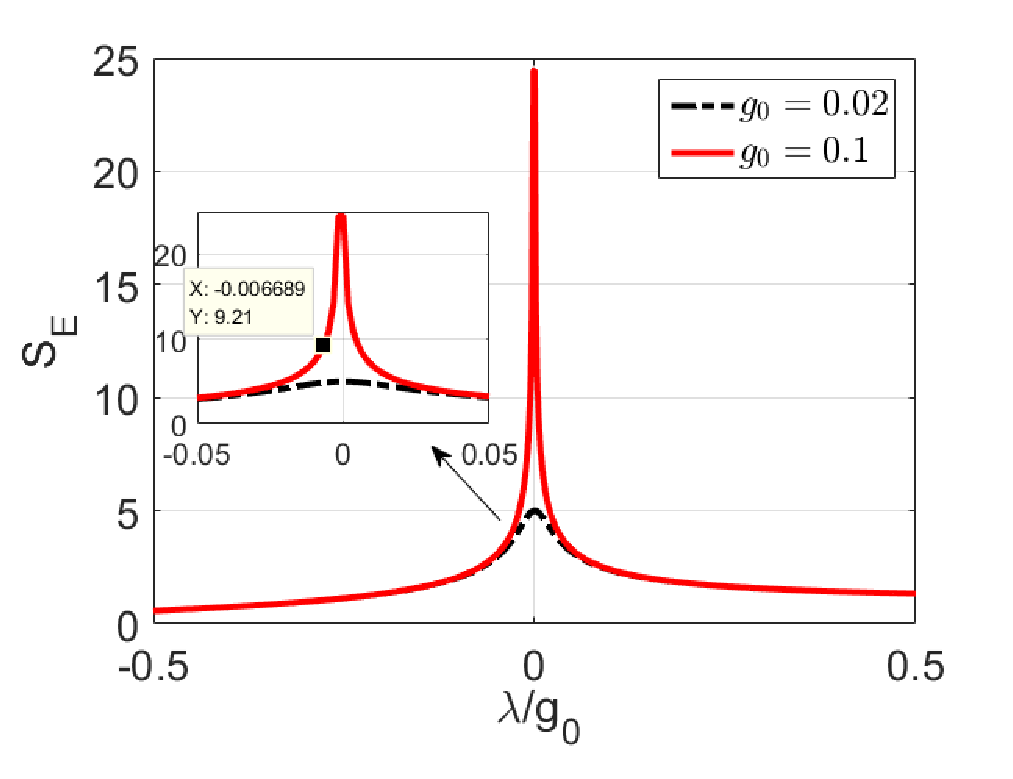}}
\caption{\label{fig-SEvsg}  The sensitivity enhancement $S_{E}$,
enabled by non-Hermiticity, as a function of $\lambda/g_{0}$ for
the NH sensor. The other parameters remain consistent with those
in Fig.~\ref{fig-Evsg}.}
\end{figure}

As illustrated in Figs.~\ref{fig-Evsg}(a) and~\ref{fig-Evsg}(b),
when $g_{0}=0.02$ and $g_{0}=0.1$, which are relatively small
values ($0 < c \leq \max[a(t) - b(t)]$), the eigenvalue-matching
condition for the NH sensor is satisfied. This results in a
minimum energy splitting of $|\Delta_E(\tau)|_{\textrm{min}}
\approx 8.03\times 10^{-3}$ for $g_0 = 0.02$ at $\tau =
0.295\pi/\omega$ and $|\Delta_E(\tau)|_{\textrm{min}} \approx
8.17\times 10^{-3}$ for $g_0 = 0.1$ at  $\tau=\pi/\omega$, during
one period. Therefore, even a smaller deviation of the interaction
strength $g$ from $g_{0}$ leads to enhanced sensitivity
$|\chi_{EP}(\tau)|$ in both Figs.~\ref{fig-Evsg}(a)
and~\ref{fig-Evsg}(b). However, the  sensitivity peak illustrated
in Fig.~\ref{fig-Evsg}(b) surpasses that in
Fig.~\ref{fig-Evsg}(a), attributed to the higher fixed parameter
value  \( g_{0} = 0.1 \) in Fig.~\ref{fig-Evsg}(b). On the other
hand, under the Hermitian mechanism, as illustrated in
Figs.~\ref{fig-Evsg}(a) and \ref{fig-Evsg}(b), the system's
sensitivity $|\chi_{H}(\tau)|$ remains nearly invariant regardless
of variations in the parameter $\lambda/g_{0}$. This is
irrespective of whether the eigenvalue-matching condition is
fulfilled. Hence, it can be concluded that, in the NH scenario,
the eigenenergy exhibits sharper variations with respect to small
changes in $\lambda/g_{0}$ compared to its Hermitian counterpart.

To intuitively compare the enhanced sensitivity achieved using EP
with that achieved using conventionally DP in quantum sensing, we
investigate the non-Hermiticity-enabled sensitivity enhancement
factor \( S_{E} \). For small values of \( g_{0} \) [e.g., \(
g_{0}=0.02 \) and \( g_{0}=0.1 \)], the sensitivity enhancement
factor \( S_{E} \) increases monotonically as \( |\lambda/g_{0}|
\) decreases. Ideally, this enhancement can reach up to 9.21 when
the deviation \( |\lambda/g_{0}| = 0.006689 \) (see the red lines
in Fig.~\ref{fig-SEvsg}). This finding aligns with the performance
of existing time-independent NH
sensors~\cite{single-spin,pseudo-hermitian1,qubit-resonator,pseudo-hermitian2},
demonstrating that NH sensors offer a significant advantage over
conventional Hermitian sensors in estimating \( \lambda \).

\section{eigenstate-based quantum sensing}\label{section:IV}

Recent studies have demonstrated that asymmetric (chiral) state
transfer can be achieved by tuning system parameters along a path
encircling an EP or an approximate exceptional point
\cite{Arkhipov-encircling2,wu-encircling,bell-encircling,Feilhauer-encircling,Ergoktas-encircling,Arkhipov-encircling1,Tang-encircling,Nasari-encircling,wu-encircling1}.
During this quantum state transformation, the eigenstate
populations of the system undergo rapid change  at a specific
time. This feature is significant because it indicates that abrupt
shifts in the population of eigenstates, in time-dependent
systems, can serve as indirect but accurate indicators of changes
in system parameters. In this work, we aim to determine the
optimal range of parameter sensitivity in time-dependent NH
sensors through numerical analysis of system population dynamics.
Additionally, we propose a theoretical framework for
eigenstate-based quantum sensing to detect subtle deviations in
the interaction strength.

\subsection{The optimal time window for achieving high sensitivity}

We first focus on the unperturbed system dynamics governed by
$H_{S}(t)$ with the sensor initialized in one of its right
eigenstates, i.e., $|\Psi(0)\rangle$=$|\phi_{+}(0)\rangle$.
Generally, the evolving state $|\Psi(t)\rangle$ can be obtained by
numerically calculating the following Schr\"{o}dinger's equation
\begin{eqnarray}\label{eq1-13}
i\partial_{t}|\Psi(t)\rangle=H_{S}(t)|\Psi(t)\rangle.
\end{eqnarray}
As shown in
Refs.~\cite{Arkhipov-encircling2,wu-encircling,bell-encircling,
Feilhauer-encircling,Ergoktas-encircling,Arkhipov-encircling1,Tang-encircling,Nasari-encircling,wu-encircling1},
efficient asymmetric or symmetric state transfers can be realized
by modulating system parameters. Specifically, an asymmetric state
transfer can be achieved via a carefully designed parameter
trajectory, such as the system time-evolution trajectory
illustrated in Eq.~(\ref{eq1-12}). To evaluate the optimal time
window for achieving high sensitivity, we investigate the temporal
dynamics of the system. The normalized population of the right
eigenstate $|\phi_{m}(t)\rangle$ ($m=+,-$) is determined by the
following relation
\begin{eqnarray}\label{eq1-14}
P_{m}(t)=\frac{|\langle\widehat{{\phi_{m}}}(t)|\Psi(t)\rangle|^2}{|\langle\widehat{{\phi_{+}}}(t)|\Psi(t)\rangle|^2+|\langle\widehat{{\phi_{-}}}(t)|\Psi(t)\rangle|^2},
\end{eqnarray}
where $|\Psi(t)\rangle$ represents the evolving state of the
system at time $t$.

\begin{figure}
\scalebox{0.45}{\includegraphics{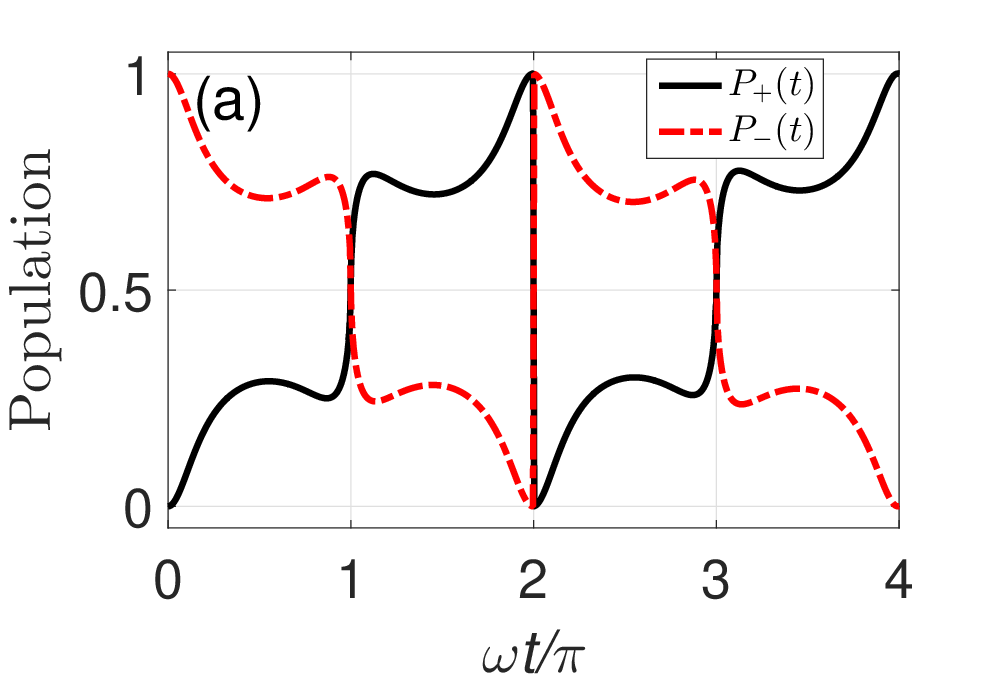}}
\scalebox{0.45}{\includegraphics{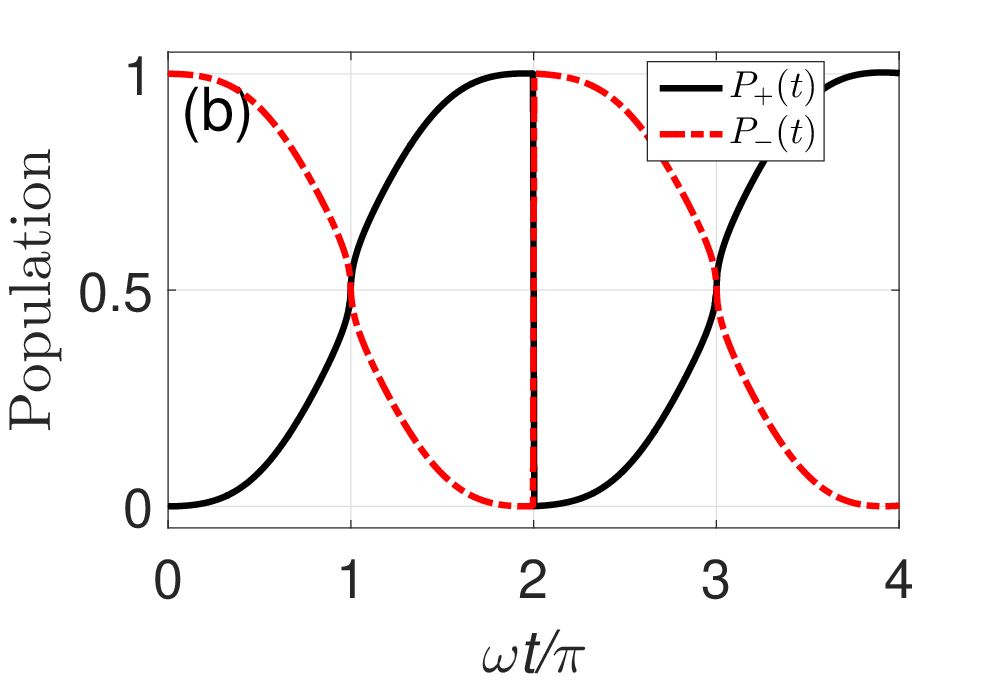}}
\caption{\label{fig-Pvst} The time evolution of the normalized
population $P_{m}(t)$ for the time-dependent right eigenstate
$|\phi_{m}(t)\rangle$ $(m=+,-)$. The initial state is chosen as
$|\Psi(0)\rangle = |\phi_{-}(0)\rangle$. In panel (a), the
parameters are set to $g_{0}=0.01$, $\Delta_{0}=0.04$,
$\Gamma_{0}=0.02$, $\omega=\pi$, and $\alpha=-1$, while in panel
(b), they are set to $g_{0}=0.02$, $\Delta_{0}=0.01$,
$\Gamma_{0}=0.04$, $\omega=\pi$, and $\alpha=-1$. The population
of the system experiences a rapid inversion when ${\omega t}/{\pi}
\approx j~(j \in \mathbf{Z}^{+})$.}
\end{figure}

In Fig.~\ref{fig-Pvst}, we illustrate the temporal evolution of
the population $P_{m}(t)$ for the right eigenstate
$|\phi_{m}(t)\rangle$ ($m=+,-$) under two sets of parameters near
the EP. Specifically, Fig.~\ref{fig-Pvst}(a) corresponds to the
parameters $g_{0}=0.01$, $\Delta_{0}=0.04$, $\Gamma_{0}=0.02$,
$\omega=\pi$, and $\alpha=-1$, while Fig.~\ref{fig-Pvst}(b) uses
$g_{0}=0.02$, $\Delta_{0}=0.01$, $\Gamma_{0}=0.04$, $\omega=\pi$,
and $\alpha=-1$. It is observed that when ${\omega t}/{\pi}
\approx j$ ($j \in \mathbf{Z}^{+}$), the  population $P_{m}(t)$
undergoes a rapid inversion. Notably, a small deviation $\Delta t$
from the fixed time $t_{0} \approx j{\pi}/\omega$ results in a
significant change in the population. Consequently, the population
signal described by Eq.~(\ref{eq1-14}) exhibits high sensitivity
to variations in time $t$. On the other hand, at the specific time
$t_{0} \approx j{\pi}/\omega$, the parameters defined in
Eq.~(\ref{eq1-12}) are fixed as follows: $g=g_{0}$,
$\delta(t_{0})=0$, and
$\gamma_{-}(t_{0})=\Gamma_{0}\sin^2(j{\pi}/2)$. A small deviation
$\Delta t$ from this fixed time $t_0$ leads to correspondingly
small deviations in these parameters. This phenomenon is
significant because it allows for the indirect but precise
determination of parameter changes by analyzing abrupt shifts in
the quantum state population of time-dependent systems.
Consequently, this system can serve as an efficient sensor for
detecting variations in the fixed parameters $[g, \delta(t_0),
\gamma_{-}(t_0)]$.

It is crucial to emphasize that two distinct physical mechanisms
are involved in the entire dynamic process. (i) The nontrivial
population conversion, characterized by rapid population
coalescence within half a period as illustrated in
Fig.~\ref{fig-Pvst}, originates from dynamically varying system
parameters along a path enclosing an EP. This phenomenon is a
unique attribute of non-Hermitian systems. (ii) The trivial
population conversion, where the state transformation occurs over
one full period, results from dynamically varying system
parameters along a path enclosing a DP, which is a common feature
of Hermitian systems. Specifically, the system functions as a NH
sensor when ${\omega t}/{\pi} \approx 2k-1$ (where $k \in
\mathbf{Z}^{+}$ and the dissipative parameter $\gamma_{-}(t_{0}) =
\Gamma_{0}$), or as a Hermitian sensor when ${\omega t}/{\pi}
\approx 2k$ (where $k \in \mathbf{Z}^{+}$ and the dissipative
parameter $\gamma_{-}(t_{0}) = 0$). Consequently, the
time-modulated quantum sensor exhibits differential sensitivities
to changes in the fixed parameters $[g(t_{0}), \delta(t_{0}),
\gamma_{-}(t_{0})]$.

Without loss of generality, we still assume that the parameter $g$
deviates slightly from its fixed value $g_{0}$, such that $\lambda
= g - g_{0}$ represents the deviation.  Then, according to
Eq.~(\ref{eq1-5}), the perturbed left eigenvectors can thus be
expressed as
\begin{eqnarray}\label{eq1-15}
\langle\widehat{\phi_{+}}(t,\lambda)|&=&\frac{1}{S_{+}(t,\lambda)}\{A_{+}(t,\lambda),4g(\lambda)\},\cr
\langle\widehat{\phi_{-}}(t,\lambda)|&=&\frac{1}{S_{-}(t,\lambda)}\{A_{-}(t,\lambda),4g(\lambda)\},
\end{eqnarray}
where
$S_{\pm}(t,\lambda)=\sqrt{|A_{\pm}(t,\lambda)|^2+|4g(\lambda)|^2}$,
$A_{\pm}(t,\lambda)=2\delta(\tau)-i[\gamma_{-}(\tau)-\gamma_{+}(\tau)]\pm{2\Delta_E(t,\lambda)}$,
and
$\Delta_E(t,\lambda)=\frac{i}{2}\sqrt{[\gamma_{-}(\tau)-\gamma_{+}(\tau)+2i\delta(\tau)]^2-16g(\lambda)^{2}}$.

The  normalized population of the right eigenstate
$|\phi_{m}(\tau)\rangle$ ($m$=$+,-$) is determined by the relation
\begin{eqnarray}\label{eq1-16}
P_{m}(t,\lambda)=\frac{|\langle\widehat{{\phi_{m}}}(t,\lambda)|\Psi(t)\rangle|^2}{|\langle\widehat{{\phi_{+}}}(t,\lambda)|\Psi(t)\rangle|^2+|\langle\widehat{{\phi_{-}}}(t,\lambda)|\Psi(t)\rangle|^2}.
\end{eqnarray}
Based on the outcomes of the measurements on the sensor system,
one can derive the susceptibility
\begin{eqnarray}\label{eq1-17}
\chi(t,\lambda)=\frac{\partial{P_+(t,\lambda)}}{\partial{\lambda}}=\frac{P_+(t,\lambda_{i+1})-P_+(t,\lambda_{i})}{\lambda_{i+1}-\lambda_{i}},
\end{eqnarray}
where the subscript $i$ denotes the data index. Eq.~(\ref{eq1-17})
quantifies the sensitivity of the population $P_{+}(t,\lambda)$ to
changes in the parameter $\lambda$ for a given evolution time $t$.
To uniformly evaluate the performance of the quantum sensor in
both Hermitian and NH regimes, we further characterize the
sensitivity enhancement enabled by non-Hermiticity as
\begin{eqnarray}\label{eq1-18}
S_{P}(t,\lambda)
=|\frac{\chi_{NH}(t,\lambda)}{\chi_{H}(t,\lambda)}|,
\end{eqnarray}
where $\chi_{NH}(t)$ and $\chi_{H}(t)$ are the susceptibility of
the quantum sensor in Hermitian and NH regimes, respectively.

\subsection{Performance of eigenstate-based quantum sensing}

\begin{figure}
\scalebox{0.4}{\includegraphics{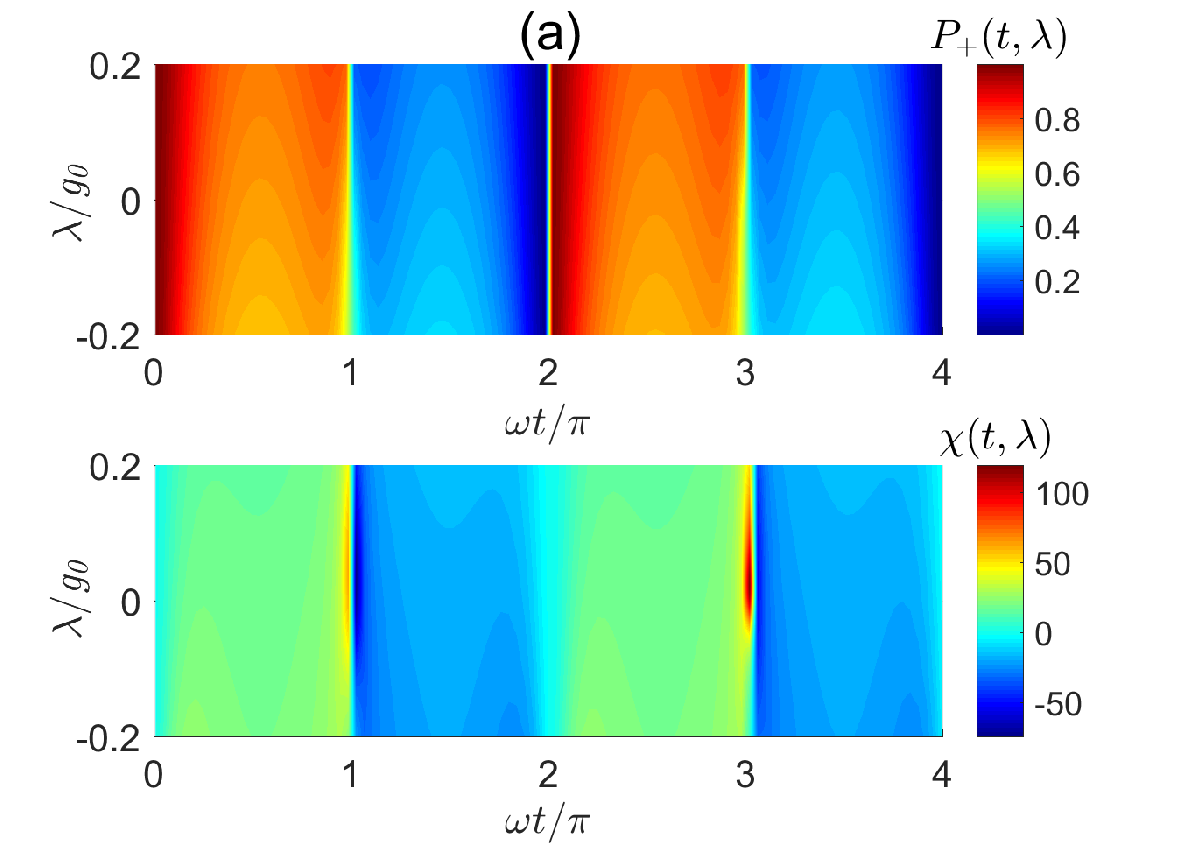}\includegraphics{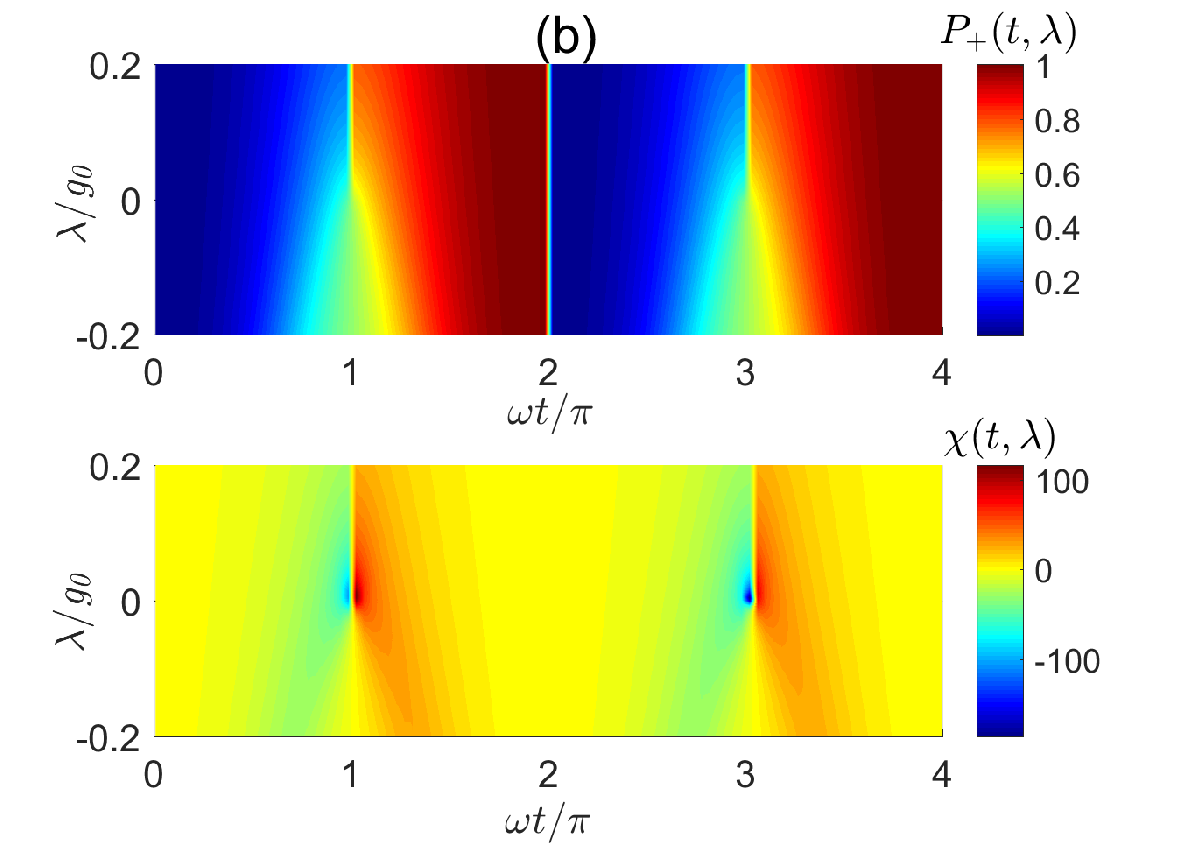}}
\caption{\label{fig-Pvstlambda}  The normalized population \(
P_{+}(t,\lambda) \) of the right eigenstate \(
|\phi_{+}(t,\lambda)\rangle \) and the susceptibility \(
\chi(t,\lambda) \) of the NH sensor as functions of \(
\lambda/g_{0} \) at different times.  In panel (a), the parameters
are set to \( g_{0}=0.01 \), \( \Delta_{0}=0.04 \), \(
\Gamma_{0}=0.02 \), \( \omega=\pi \), and \( \alpha=-1 \), while
in panel (b), they are set to \( g_{0}=0.02 \), \( \Delta_{0}=0.01
\), \( \Gamma_{0}=0.04 \), \( \omega=\pi \), and \( \alpha=-1 \).
The susceptibility \( \chi(t,\lambda) \) exhibits extremely sharp
peaks  near \({\omega t}/{\pi} \approx 2k-1 \) where \( k \in
\mathbf{Z}^{+} \).}
\end{figure}

In section~\ref{section:IV}A, we propose that the phenomenon of
rapid inversion of the eigenstate population when ${\omega
t}/{\pi} \approx j~(j \in \mathbf{Z}^{+})$, as illustrated in
Fig.~\ref{fig-Pvst}, can be harnessed to achieve high sensitivity
in quantum sensing. Moreover, the sensitivities of the sensors
exhibit distinct behaviors: for the NH sensor, the sensitivity
peaks when ${\omega t}/{\pi} \approx 2k-1$, while for the
Hermitian sensor, it peaks when ${\omega t}/{\pi} \approx 2k$. To
more rigorously validate this hypothesis, we examine the behavior
of the normalized population $P_{+}(\tau)$ of the right eigenstate
$|\phi_{+}(\tau)\rangle$ and the susceptibility $\chi(\tau)$ of
the NH sensor, as shown in Fig.~\ref{fig-Pvstlambda}. Here, we use
two sets of parameters near the EP, which have been discussed in
Fig.~\ref{fig-Pvst}.

One can observe that when $\lambda/g_{0}$ is fixed at a value such
as 0.1, the normalized population $P_{+}(t,\lambda)$ in
Figs.~\ref{fig-Pvstlambda}(a) and \ref{fig-Pvstlambda}(b)
undergoes rapid inversions near the times ${\omega t}/{\pi}
\approx 1, 2, 3$. Additionally, for the times ${\omega t}/{\pi}
\approx 1, 3$, the normalized population $P_{+}(t,\lambda)$
exhibits sharp changes near $\lambda/g_{0} = 0$. Correspondingly,
the sensitivity $\chi(t,\lambda)$ also shows significant
variations near $\lambda/g_{0} = 0$ at these times. However, at
${\omega t}/{\pi} \approx 2$, which is expected to enable
high-sensitivity detection, this expectation is not met. Despite
the rapid inversion of $P_{+}(t,\lambda)$ at ${\omega t}/{\pi}
\approx 2$, the sensitivity $\chi(t,\lambda)$ remains very small
and nearly invariant with respect to $\lambda/g_{0}$.
Consequently, for a NH sensor, the susceptibility signal described
by Eq.~(\ref{eq1-18}) exhibits high sensitivity to changes in
$\lambda/g_{0}$ only when ${\omega t}/{\pi} \approx 2k-1$ for $k
\in \mathbf{Z}^{+}$.

\begin{figure}
\scalebox{0.5}{\includegraphics{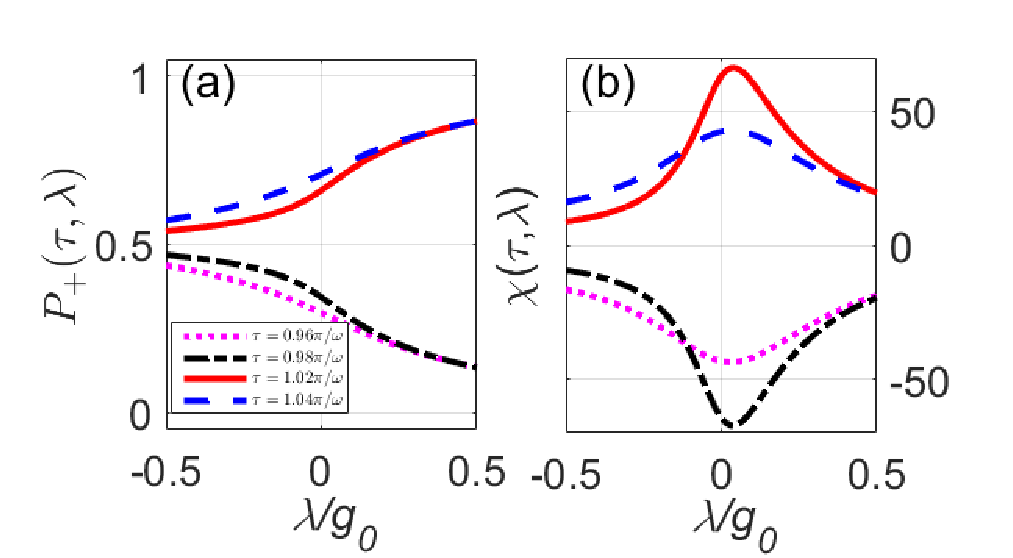}}
\scalebox{0.5}{\includegraphics{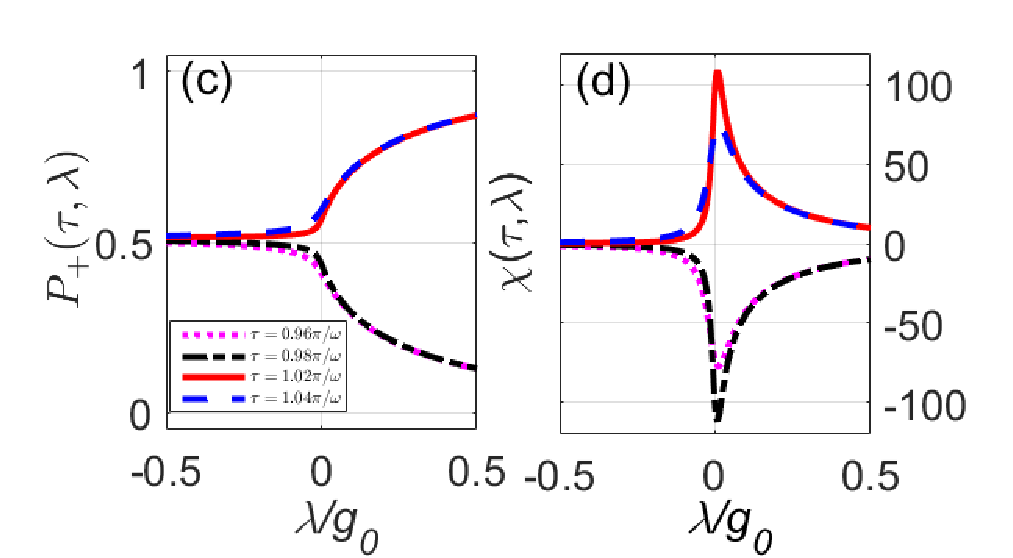}}
\caption{\label{fig-PXvsg}  The normalized population \(
P_{+}(\tau) \) of the right eigenstate \( |\phi_{+}(\tau)\rangle
\) and the susceptibility \( \chi(\tau) \) of the NH sensor as
functions of \( \lambda/g_{0} \) at different times.  In panels
(a) and (b), the parameters are set to \( g_{0}=0.01 \), \(
\Delta_{0}=0.04 \), \( \Gamma_{0}=0.02 \), \( \omega=\pi \), and
\( \alpha=-1 \), while in panels (c) and (d), the parameters are
set to \( g_{0}=0.02 \), \( \Delta_{0}=0.01 \), \( \Gamma_{0}=0.04
\), \( \omega=\pi \), and \( \alpha=-1 \). }
\end{figure}

To gain an intuitive understanding of the behavior of the
normalized population $P_{+}(\tau)$  and the susceptibility
$\chi(\tau)$ at several time points near $t \approx
{\pi}/{\omega}$
($\tau={0.96\pi}/{\omega},{0.98\pi}/{\omega},{1.02\pi}/{\omega},{1.04\pi}/{\omega}$),
we present the results in Fig.~\ref{fig-PXvsg}. As illustrated in
Fig.~\ref{fig-PXvsg}(a), the normalized population $P_{+}(\tau)$
exhibits rapid fluctuations across all selected time nodes as
$\lambda/g_{0}$ varies. Correspondingly, in
Fig.~\ref{fig-PXvsg}(b), the magnitude of the susceptibility
$|\chi(\tau)|$ increases sharply with decreasing
$|\lambda/g_{0}|$. By comparing the sensitivity variations at
different time nodes, it is evident that a smaller time offset
leads to a higher peak value of $|\chi(\tau)|$. Similar trends are
observed in Figs.~\ref{fig-PXvsg}(c) and~\ref{fig-PXvsg}(d).
Notably, when comparing Fig.~\ref{fig-PXvsg}(b) and
Fig.~\ref{fig-PXvsg}(d), it is found that under identical
measurement times, the peak value of $|\chi(\tau)|$ in
Fig.~\ref{fig-PXvsg}(d) is higher than in Fig.~\ref{fig-PXvsg}(b),
albeit with slightly reduced broadening. These observations
substantiate the feasibility and efficiency of our proposed
quantum sensor based on population dynamics.

\begin{figure}
\scalebox{0.5}{\includegraphics{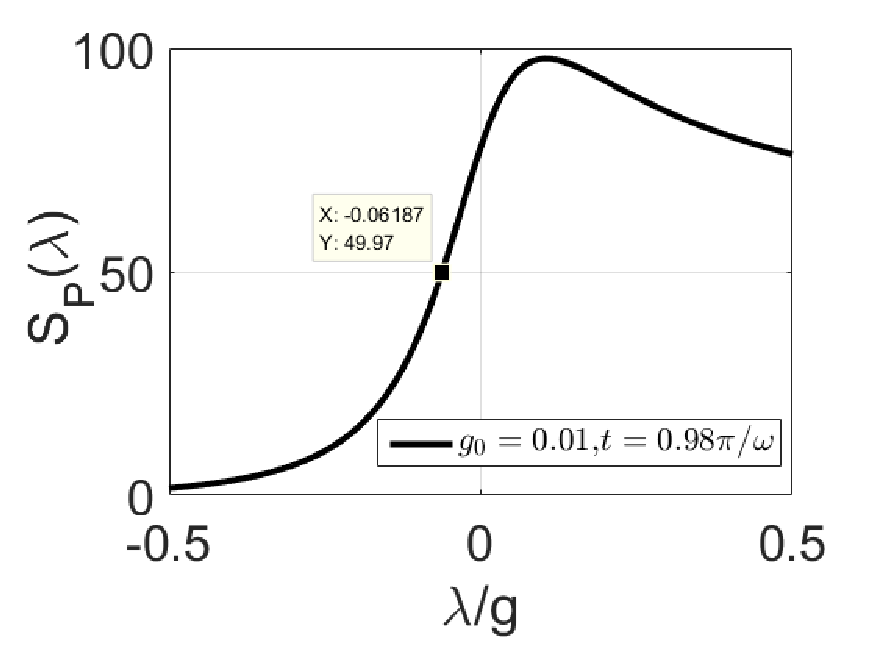}}
\caption{\label{fig-SPvstlamada} The sensitivity enhancement
enabled by non-Hermiticity of the sensor is investigated as a
function of $\lambda/g_{0}$. For the Hermitian (NH) sensor, the
dissipation factor is chosen as $\Gamma_{0}=0$
($\Gamma_{0}=0.02$). Other parameters are set as follows:
$t={0.98\pi}/{\omega}$, $g_{0}=0.01$, $\Delta_{0}=0.04$,
$\omega=\pi$, and $\alpha=-1$.}
\end{figure}

We present the non-Hermiticity-enabled sensitivity enhancement
factor $S_{P}$ as a function of $\lambda/g_{0}$ and $t$ in
Fig.~\ref{fig-SPvstlamada}. It is observed that the sensitivity
enhancement factor \( |S_{P}| \) is significantly large and
increases approximately monotonically as \( |\lambda/g_{0}| \)
decreases under both sets of parameter conditions. Specifically,
for the parameter condition with \( g_{0} = 0.01 \), the peak
value of \( S_{P} \) reaches approximately 100. Moreover, \( S_{P}
\) remains greater than 50 when \( \lambda/g_{0} > -0.06187 \),
surpassing the performance of conventional Hermitian sensors and
existing time-independent NH
sensors~\cite{single-spin,pseudo-hermitian1,qubit-resonator,pseudo-hermitian2}.

\section{Feasibility of the experimental realization}\label{section:V}

Our proposal can be realized in various physical settings, which
necessitates the engineering of the single-qubit Hamiltonian
described by Eq.~(\ref{eq-Hs}). For instance, it can be
implemented in a system of two coupled dissipative
cavities~\cite{Arkhipov-encircling2,experiment-realize1,experiment-realize2,experiment-realize3},
where $\gamma_{+}(t)$  [$\gamma_{-}(t)$] denotes the cavity gain
(loss) rate and $\delta(t)$ is the frequency detuning of the
cavities. The cavities are coupled coherently with interaction
strength $g$, while $\lambda$ accounts for deviations of the
interaction strength $g$ due to a weak external field.
Specifically, for a whispering-gallery microcavity, such as a
micro-disk or micro-toroid, under the two-mode approximation of
counter-traveling waves and perturbed by \textit{N} Rayleigh
scatterers
\cite{experiment-realize1,experiment-realize2,experiment-realize3},
the effective Hamiltonian in the traveling-wave basis
(counterclockwise and clockwise ) is
\begin{eqnarray}\label{eq3-11}
H^{(N)}=\left(\begin{array}{ccccccc}
\omega^{(N)} &  A^{(N)}\\
B^{(N)} & \omega^{(N)}  \\
\end{array}\right),
\end{eqnarray}
where $\omega^{(N)}$=$\omega_{0}$+$\sum^{N}_{j=1}\varepsilon_{j}$,
$A^{(N)}$=$\sum^{N}_{j=1}\varepsilon_{j}e^{-i2m\beta_{j}}$ and
$B^{(N)}$=$\sum^{N}_{j=1}\varepsilon_{j}e^{i2m\beta_{j}}$. Here,
 $m$ is the azimuthal mode number, $\omega_{0}$ is the complex
frequency of the unperturbed resonance mode, $\beta_{j}$ is the
angular position of scatterer $j$, and $\varepsilon_{j}$ is the
complex frequency splitting that is introduced by scatterer $j$
alone. It is possible for $|A^{(M)}|$ to differ from $|B^{(M)}|$,
due to the asymmetry in backscattering between clockwise- and
anticlockwise-travelling waves. Notably, the above Hamiltonian is
still non-Hermitian because $\varepsilon_{j}$ and $\omega_{0}$ are
complex numbers that reflect the presence of
losses~\cite{experiment-realize1}. The Hamiltonian  in
Eq.~(\ref{eq-Hs}) can also be realized in other physical systems,
such as solid-state spin  in diamond and superconducting qubit for
magnetic field and microwave detection. In particular, the
implementation of NH dynamics has been achieved using a
nitrogen-vacancy center in diamond through selective microwave
pulses~\cite{experiment-realize4}. In addition, it has been
achieved in  a single-photon interferometric
network~\cite{experiment-realize5}. This further facilitates the
experimental realization of the present proposal.

Moreover, the feasibility of obtaining key information through
actual quantum measurements is a critical factor for enhancing the
sensitivity of quantum sensing in experimental contexts. For
eigenvalue-based quantum sensors, the sensitivity of quantum
sensing relies on the precise measurement of the energy splitting
$\Delta_{E}$. Fortunately, recent experiments have demonstrated
that the real part of the energy splitting
$\textrm{Re}(\Delta_{E})$ corresponds to the conventional
frequency splitting, while the imaginary part
$\textrm{Im}(\Delta_{E})$ determines the linewidth. Both frequency
splitting and linewidth can be accurately measured experimentally
in a whispering-gallery microcavity (for details, see
Refs.~\cite{quantum-sensing4,Microcavity-sensor,experiment-realize1,exceptional_surfaces,linewidth,EP_shift}).
Furthermore, from an experimental perspective, based on the
specific experimental parameters, precisely measuring either the
frequency splitting or the linewidth is sufficient to achieve
enhanced sensitivity. For instance, as shown in
Fig.~\ref{fig-Evstg}(a), although the sensor sensitivities
obtained by measuring the real and imaginary parts are not
significantly different, it is evident that the imaginary part
$\textrm{Im}(\Delta_{E})$ is substantially larger than its real
part $\textrm{Re}(\Delta_{E})$. In this case, choosing to measure
the imaginary part, i.e., the linewidth of the system, is more
practical and reasonable experimentally. Similarly, under the
system parameters illustrated in Fig.~\ref{fig-Evstg}(b),
selecting to measure the frequency splitting in the system is more
advisable for achieving higher sensor sensitivity.

On the other hand, for the eigenstate-based quantum sensor, the
sensitivity of quantum sensing depends on the precise measurement
of the population of the right eigenstate $P_{m}(t,\lambda)$ ($m =
+, -$). It should be noted that accurately measuring the
population of the eigenstate at the EP poses significant
experimental challenges. This is because the eigenstates of
non-Hermitian systems become identical at the EP, and due to their
indistinguishability in non-Hermitian systems, determining the
population of the eigenstate becomes difficult. Fortunately, the
technique for quantum state discrimination has been extensively
studied~\cite{discriminating1,discriminating2} and has already
been experimentally realized in recent studies within
non-Hermitian
systems~\cite{experiment-discriminating1,experiment-discriminating2}.
Moreover, one can also choose to appropriately deviate the system
parameter trajectory from the EP, thereby enhancing the
distinguishability between eigenstates and alleviating the
experimental challenge associated with measuring the population of
eigenstates. In fact, as shown in the Appendix~\ref{appendixA},
the eigenstate-based sensor still exhibits significant
susceptibility during dynamical evolution, even when  not close to
an EP.

\section{CONCLUSION}\label{section:VI}

We have proposed two theoretical schemes to achieve enhanced
quantum sensing in time-modulated non-Hermitian systems by
leveraging the coalescence of eigenvalues and eigenstates. This
approach contrasts sharply with previous studies, where the
Hamiltonians of the designed sensors were time-independent. We
have conducted a thorough investigation into various features
associated with optimal conditions for sensitivity enhancement,
such as the optimal timing window for achieving high sensitivity
using the energy spectrum of eigenvalues and the population
distribution of eigenstates. Notably, we have demonstrated that
when the  time-evolution trajectory of the system approaches the
EP, both eigenvalue-based and eigenstate-based quantum sensors
exhibit significant enhancements in sensitivity, 9.21-fold and
50-fold, respectively, compared to conventional Hermitian sensors.
Moreover, we emphasize that the eigenstate-based sensor still
demonstrates divergent susceptibility even when the system's
time-evolution trajectory deviates from the EP (for details, see
Appendix~\ref{appendixA}). However, there is a trade-off between
achieving high sensitivity (through a small perturbation
$\lambda$) and ensuring the reliability of quantum sensing
measurements (with a low uncertainty ratio
${\Delta}\lambda'/\lambda$ for the perturbation), especially when
considering the effects of background noise and the limitations of
practical measurements (for more details, see
Appendices~\ref{appendixB} and \ref{appendixC}). Our findings pave
a novel pathway for advanced sensing in a time-sensitive
framework, thereby complementing ongoing efforts to leverage the
distinctive properties of open systems.

The presented results are not only applicable to single-qubit NH
systems, but also can be extended to time-modulated multi-mode NH
systems that dynamically encircle an EP or
DP~\cite{Arkhipov-encircling1,bell-encircling}. In this context,
multi-mode systems with higher-order EPs or multiple low-order EPs
may exhibit more complex and novel phenomena. Moreover, extending
this work to explore sensors in time-dependent open quantum
systems, based on the exceptional points within the
hybrid-Liouvillian formalism~\cite{Liouvillians,Liouvillians1},
represents a promising direction for future research.
Additionally, by considering applications in multiple sensing
scenarios over a time sequence and integrating other enhancement
methods, such as exceptional surfaces~\cite{exceptional_surfaces},
the EP shift sensing mechanism~\cite{EP_shift}, and the linewidth
broadening mechanism~\cite{linewidth}, the performance of sensors
can be further enhanced, which also represents a promising
direction for future research.

\section*{ACKNOWLEDGEMENT}
This work was supported by National Key Research and Development
Program of China (Grant No. 2024YFA1408900), National Natural
Science Foundation of China (NSFC) (Grants Nos. 12264040,
12374333, 12364048 and U21A20436),  Jiangxi Natural Science
Foundation (Grant Nos. 20232BCJ23022 and 20242BAB25037),
Innovation Program for Quantum Science and Technology (Grant
No.~2021ZD0301705), and Jiangxi Province Key Laboratory of Applied
Optical Technology (Grant No.~2024SSY03051).

\appendix

\section{The impact of distancing from an exceptional point} \label{appendixA}

\begin{figure*}
\scalebox{0.45}{\includegraphics{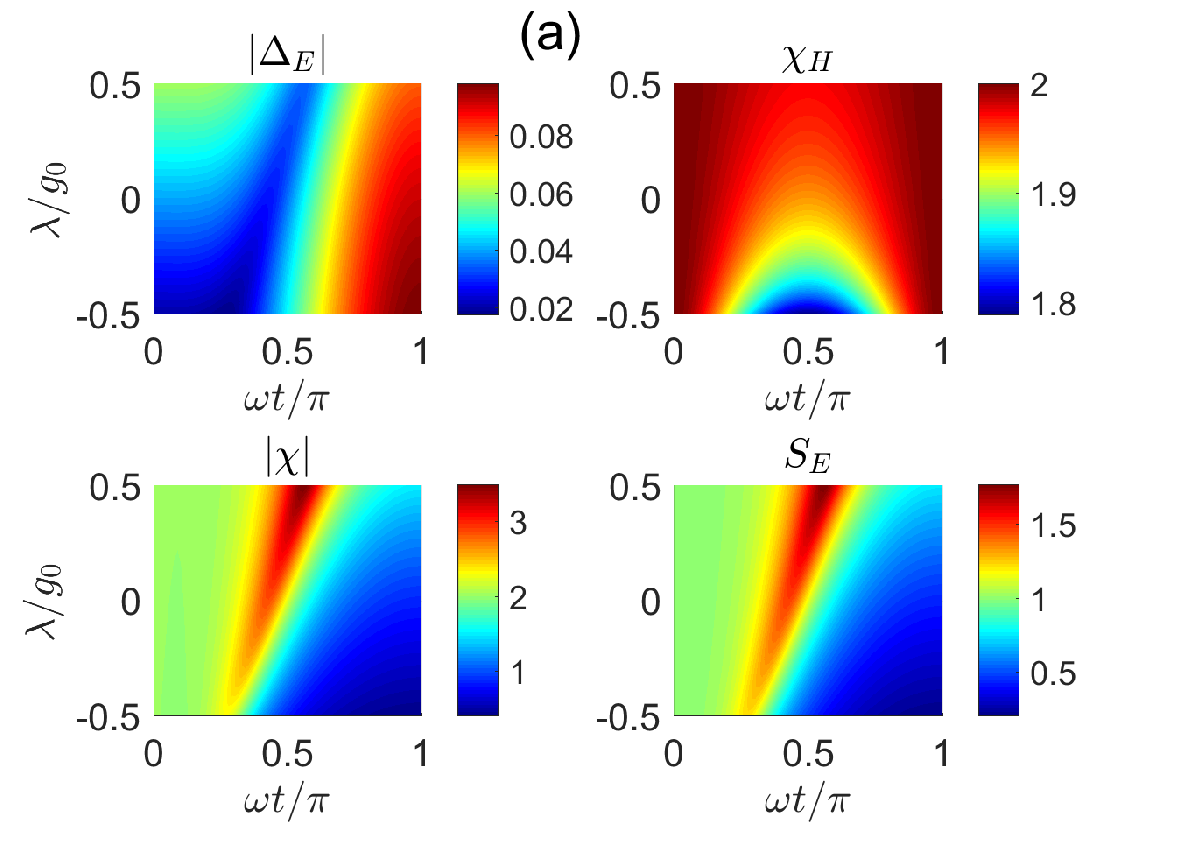}\includegraphics{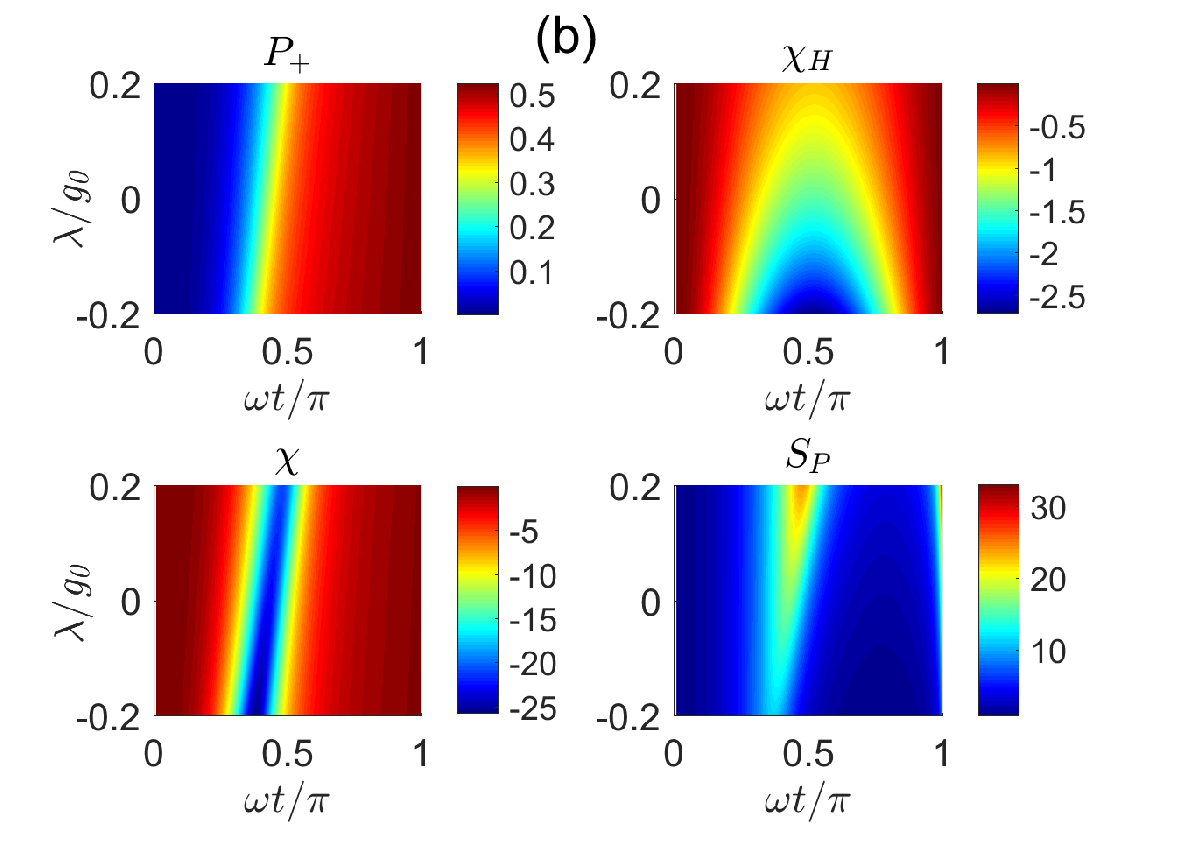}}
\caption{\label{fig-EfarfromEP} (a) The parameters $|\Delta_{E}|$,
$|\chi|$, $\chi_H$, and $S_E$ are analyzed as functions of
$\lambda/g_{0}$ and $t$ for the eigenvalue-based quantum sensor.
(b) For the eigenstate-based quantum sensor, the parameters
$P_{+}$, $\chi$, $\chi_H$, and $S_P$ are analyzed as functions of
$\lambda/g_{0}$ and $t$. Other parameters are set to $g_{0}=0.02$,
$\Delta_{0}=0.04$, $\Gamma_{0}=0.1$, $\omega=\pi$, and
$\alpha=-1$.}
\end{figure*}

In the main text, we have examined the performance of the quantum
sensor near the EP. Here, we briefly analyze the performance of
the scheme when the time-evolution trajectory of the system is
situated away from the EP. Without loss of generality, the
time-evolution trajectory of the system follows that depicted by
Eq.~(\ref{eq1-12}), with the parameters set as $g_{0}=0.02$,
$\Delta_{0}=0.04$, $\Gamma_{0}=0.1$, $\omega=\pi$, and
$\alpha=-1$. Additionally, as discussed in Section
~\ref{section:III}A, the parameter
$|\Delta_{E}(\tau)|_{\textrm{min}}$ serves as an indicator used to
evaluate the extent to which the trajectory of the system
parameters deviates from the EP or DP.

For the eigenvalue-based quantum sensor, as indicated in
Eq.~(\ref{eq1-6}), the susceptibility of the quantum sensor is
contingent upon the eigenvalue-matching condition ($\Delta_E(\tau)
\rightarrow 0$ near the EP or DP). By comparing
Figs.~\ref{fig-Evsg} and \ref{fig-SEvsg} with
Fig.~\ref{fig-EfarfromEP}(a), it is evident that both the
susceptibility $|\chi|$ and the sensitivity enhancement factor
$S_{E}$ of the quantum sensor diminish significantly when the
time-evolution trajectory of the system is distant from the EP,
specifically when $|\Delta_{E}(\tau)|_{\textrm{min}} > 0.02$.
However, by comparing Figs.~\ref{fig-Pvstlambda} and
\ref{fig-SPvstlamada} with Fig.~\ref{fig-EfarfromEP}(b), it
becomes clear that for the eigenstate-based quantum sensor, a
relatively high susceptibility
$|\chi_H(t,\lambda)|_{\textrm{max}}=25$ is still observed.
Moreover, the non-Hermiticity-enhanced susceptibility $S_P$
exceeds 20 over a wide range of parameters. Consequently, the
eigenstate-based sensor exhibits significant susceptibility during
dynamical evolution, even when not close to an EP.

\section{Quantum sensing under noise} \label{appendixB}

Noise represents a fundamental challenge for sensors, as the
performance of a sensor is ultimately constrained by
noise~\cite{quantum-sensing2}. Without loss of generality, we now
proceed to analyze the performance of the eigenstate-based sensor
while taking into account both background noise and the
imperfections inherent in actual measurements.

Under the combined effects of background noise and the
imperfections in actual measurements, the measured population of
the right eigenstate in Eq.~(\ref{eq1-16}) is accordingly modified
as~\cite{quantum-sensing2}
\begin{eqnarray}\label{eqa-1}
P'_{+}(t,\lambda)&=&\frac{N_{+}(t,\lambda)+N'_{+}(t,\lambda)}{N_{+}(t,\lambda)+N'_{+}(t,\lambda)+N_{-}(t,\lambda)+N'_{-}(t,\lambda)}
\cr&&{=}\frac{N_{+}(t,\lambda)}{N_{+}(t,\lambda)+N_{-}(t,\lambda)}\frac{1+\frac{N'_{+}(t,\lambda)}{N_{+}(t,\lambda)}}{1+\frac{N'_{+}(t,\lambda)+N'_{-}(t,\lambda)}{N_{+}(t,\lambda)+N_{-}(t,\lambda)}}
\cr&&{\approx}\frac{N_{+}(t,\lambda)}{N_{+}(t,\lambda)+N_{-}(t,\lambda)}[1+\frac{N'_{+}(t,\lambda)}{N_{+}(t,\lambda)}][1-\frac{N'_{+}(t,\lambda)+N'_{-}(t,\lambda)}{N_{+}(t,\lambda)+N_{-}(t,\lambda)}]
\cr&&{\approx}\frac{N_{+}(t,\lambda)}{N_{+}(t,\lambda)+N_{-}(t,\lambda)}[1+\frac{N'_{+}(t,\lambda)}{N_{+}(t,\lambda)}-\frac{N'_{+}(t,\lambda)+N'_{-}(t,\lambda)}{N_{+}(t,\lambda)+N_{-}(t,\lambda)}]
\cr&&=P_{+}(t,\lambda)+[1-P_{+}(t,\lambda)]\frac{N'_{+}(t,\lambda)}{N}-P_{+}(t,\lambda)\frac{N'_{-}(t,\lambda)}{N},
\end{eqnarray}
where
$N_{m}(t,\lambda)=|\langle\widehat{{\phi_{m}}}(t,\lambda)|\Psi(t)\rangle|^2$
($m=+,-$), $N'_{m}(t,\lambda)$ quantify the additional noisy
contributions to the corresponding eigenstate populations caused
by background noise, and $N=N_{+}(t,\lambda)+N_{-}(t,\lambda)$.
Here, we assume that
$\frac{1}{1+\frac{N'_{+}(t,\lambda)+N'_{-}(t,\lambda)}{N_{+}(t,\lambda)+N_{-}(t,\lambda)}}
\approx
1-\frac{N'_{+}(t,\lambda)+N'_{-}(t,\lambda)}{N_{+}(t,\lambda)+N_{-}(t,\lambda)}$
when
$\frac{N'_{+}(t,\lambda)+N'_{-}(t,\lambda)}{N_{+}(t,\lambda)+N_{-}(t,\lambda)}
\ll 1$. In general, background noise is completely random. The
corresponding population results are assumed to be uniformly
distributed within the intervals $N'_{+}(t,\lambda)\in[0,
\eta_{+}N]$ and $N'_{-}(t,\lambda)\in[0, \eta_{-}N]$,
respectively. However, one can fix the average values of
$\eta_{+}$ and $\eta_{-}$ by performing repetitive measurements
with the same apparatus.

Then, by applying the error propagation procedure to
Eq.~(\ref{eqa-1}), the measurement noise of $P'_{+}(t,\lambda)$
can be determined as
\begin{eqnarray}\label{eqa-2}
{\Delta}P^{'2}_{+}(t,\lambda)&=&[\frac{\partial{P'_{+}(t,\lambda)}}{{\partial}P_{+}(t,\lambda)}\Delta{P_{+}(t,\lambda)}]^2
+[\frac{\partial{P'_{+}(t,\lambda)}}{{\partial}N'_{+}(t,\lambda)/N}\Delta{\frac{N'_{+}(t,\lambda)}{N}}]^2
+[\frac{\partial{P'_{+}(t,\lambda)}}{{\partial}N'_{-}(t,\lambda)/N}\Delta{\frac{N'_{-}(t,\lambda)}{N}}]^2\cr&&
=[1-({\eta_{+}+\eta_{-}})]^2{\Delta}P^{2}_{+}(t,\lambda)
+\{[1-P_{+}(t,\lambda)]^2\eta^{2}_{+}+P^{2}_{+}(t,\lambda)\eta^{2}_{-}\},
\end{eqnarray}
where
${\Delta}P^{2}_{+}(t,\lambda)=\frac{P^{2}_{+}(t,\lambda){\Delta}P^{2}_{-}(t,\lambda)+P^{2}_{-}(t,\lambda){\Delta}P^{2}_{+}(t,\lambda)}{N^{4}}$
is the standard deviation for the quantum sensor, arising from the
fluctuations in population measurements and can be averaged out
through repetitive measurements. Therefore, the first term on the
right-hand side of Eq.~(\ref{eqa-2}) may tend to zero with
sufficiently many repetitive measurements. However, the second
term is independent of the fluctuations in population measurements
and cannot be averaged out, as it originates from background
noise. Based on the above analysis, we can approximate the minimum
measurement uncertainty of the estimated parameter as follows:
\begin{eqnarray}\label{eqa-4}
{\Delta}\lambda'&=&\frac{{\Delta}P^{'}_{+}(t,\lambda)}{|{\partial}_{\lambda'}P^{'}_{+}(t,\lambda)|}
\leq\frac{{\Delta}P^{'}_{+}(t,\lambda)}{|[1-({\eta_{+}+\eta_{-}})]{\partial}_{\lambda}P^{}_{+}(t,\lambda)|}
{\geq}\frac{{\Delta}P^{'}_{+}(t,\lambda)}{|\chi(t,\lambda|},
\end{eqnarray}
with $({\eta_{+}+\eta_{-}})\ll1$, i.e.,
$\textrm{Min}[{\Delta}\lambda']\approx{{\Delta}P^{'}_{+}(t,\lambda)}/{|\chi(t,\lambda|}$.

\begin{figure}
\scalebox{0.47}{\includegraphics{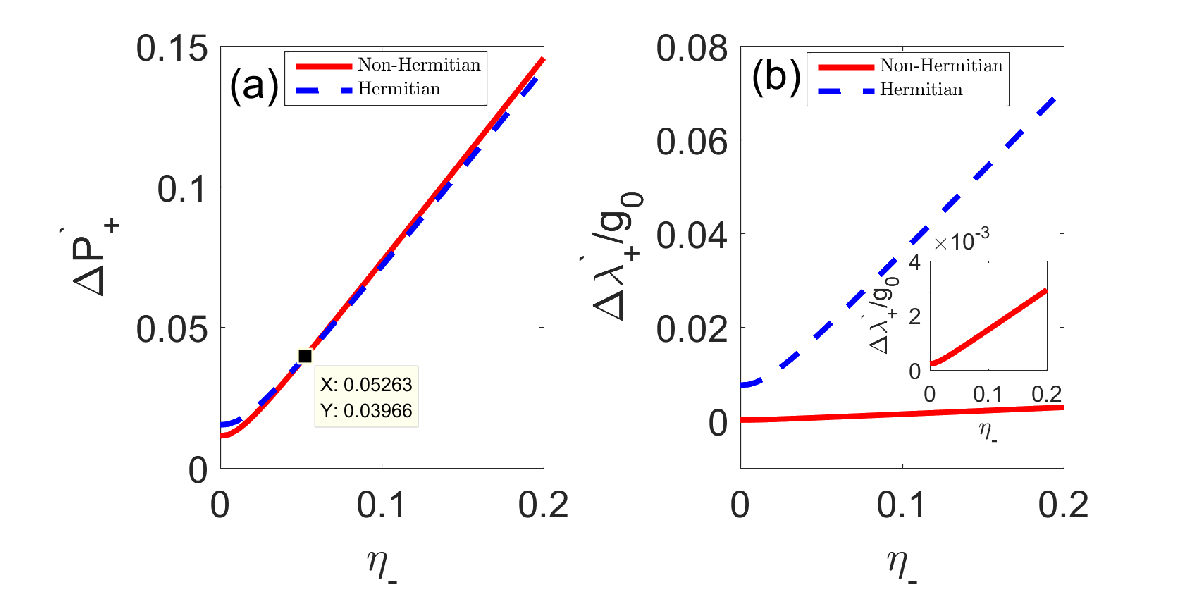}}
\caption{\label{fig-lambda_P} (a) The standard deviation of
${\Delta}P^{'}_{+}$ as a function of $\eta_{-}$. (b) The standard
deviation of ${\Delta}\lambda'$ as a function of $\eta_{-}$. The
red curves denote the non-Hermitian sensor, while the blue curves
represent the results of its Hermitian counterpart. We choose the
working point of the sensor in Fig.~\ref{fig-SPvstlamada}, i.e.,
$\lambda/g_{0}=-0.618$, $S_{P}=49.97$ when $t={0.98\pi}/{\omega}$.
The the population $P_{+}=0.3784$ for the non-Hermitian sensor and
$P_{+}=0.49$ for its Hermitian counterpart. Other parameters are
set to $\eta_{+}=\eta_{-}$, ${\Delta}P_{+}=P_{+}{\times}3\%$.}
\end{figure}

Without loss of generality, we choose the working point of the
sensor in Fig.~\ref{fig-SPvstlamada}, i.e.,
$\lambda/g_{0}=-0.618$, $S_{P}=49.97$ when $t={0.98\pi}/{\omega}$.
Then, in Fig.~\ref{fig-lambda_P}, we illustrate the dependence of
the measurement uncertainties ${\Delta}P^{'}_{+}$ and
${\Delta}\lambda'$ on the noise strength $\eta_{-}$ (assuming for
simplicity that $\eta_{+}=\eta_{-}$) for both non-Hermitian and
Hermitian sensors. The results reveal that for
$\eta_{-}\leq{0.05263}$, the uncertainty ${\Delta}P^{'}_{+}$ of
the non-Hermitian sensor is slightly lower than that of the
Hermitian sensor. Conversely, when $\eta_{-}\geq{0.05263}$, the
uncertainty ${\Delta}P^{'}_{+}$ of the non-Hermitian sensor
becomes slightly higher than that of the Hermitian sensor.
Nevertheless, the minimum value of ${\Delta}\lambda'$ for the
non-Hermitian sensor consistently remains smaller than that of the
Hermitian sensor. Consequently, the non-Hermitian sensor
demonstrates superior performance relative to its Hermitian
counterpart, achieving reduced measurement uncertainties in both
the presence and absence of background noise.

Moreover, ideally, the enhancement of the sensor can reach an
extremely high value when the perturbation $\lambda$ becomes
exceedingly small at a specific time, as illustrated in
Figs.~\ref{fig-Pvstlambda} and~\ref{fig-SPvstlamada}. In practice,
however, this theoretical prediction is not feasible in actual
experimental measurements. Specifically, as shown in
Fig.~\ref{fig-lambda_P} (b), the magnitude of the uncertainty
${\Delta}\lambda'$ is on the order of $10^{-3}g_{0}$, even with a
relatively large background noise intensity
($\eta_{+}=\eta_{-}=0.05$). This suggests that the true value of
the parameter $\lambda$ selected for measurement should ideally be
around $10^{-2}g_{0}$; otherwise, the excessively large
uncertainty ${\Delta}\lambda'$ may compromise the physical
significance of the measurement. Of course, the uncertainty of
${\Delta}\lambda'$ could be further reduced if the intensity of
background noise were decreased further. However, in some
experimental scenarios, achieving very low background noise
intensity remains a significant challenge. Therefore, there exists
a trade-off between achieving high sensitivity (via a small
perturbation $\lambda$) and ensuring the reliability of quantum
sensing measurements (with a low uncertainty ratio
${\Delta}\lambda'/\lambda$ for the perturbation), particularly
when accounting for the impact of background noise and the
constraints of practical measurements.

\section{Quantum sensing based on the master equation} \label{appendixC}

\begin{figure}
\scalebox{0.4}{\includegraphics{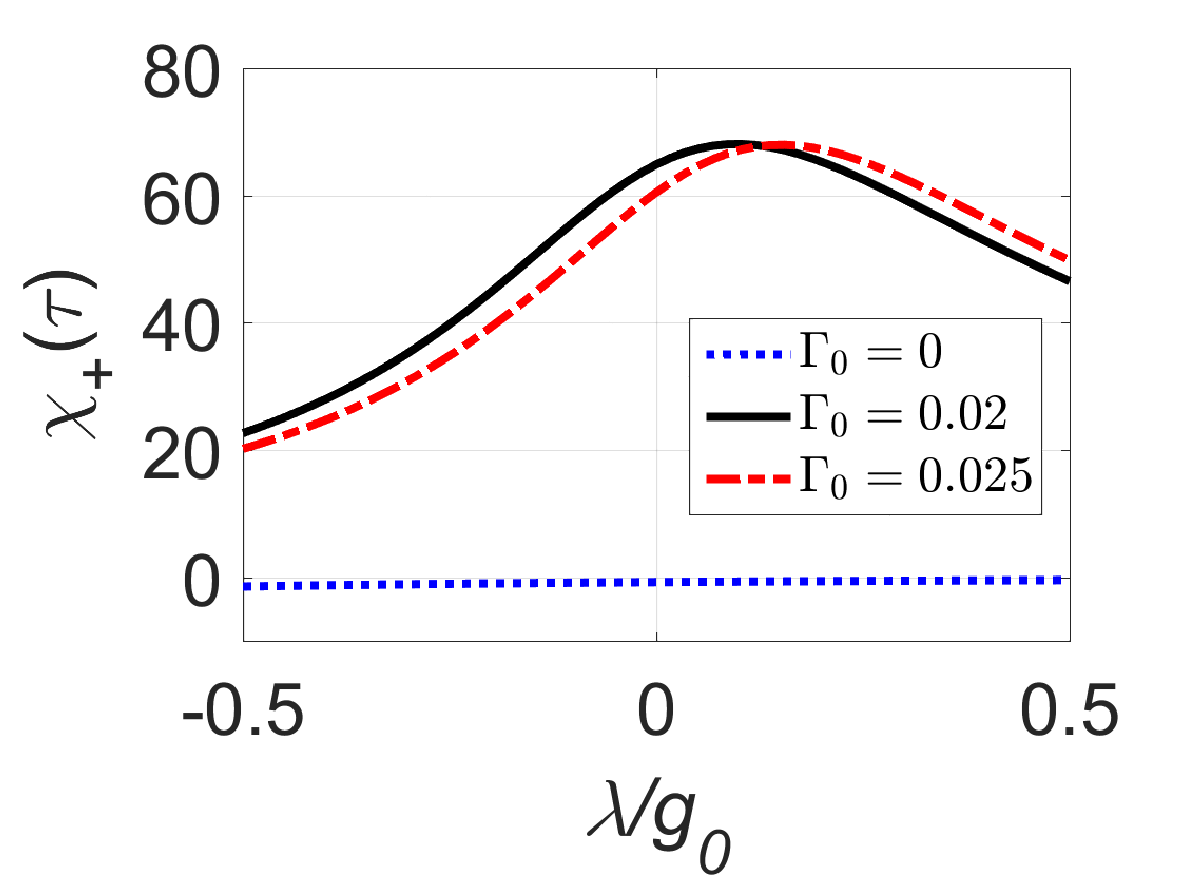}}
\caption{\label{fig-Appendix_C} The susceptibility \( \chi(\tau)
\) of the quantum sensing based on the master equation as
functions of \( \lambda/g_{0} \) under different dissipation
coefficients $\Gamma_{0}$.   Other parameters are set to
$\tau={0.98\pi}/{\omega}$, \( g_{0}=0.01 \), \( \Delta_{0}=0.04
\), and \( \alpha=-1 \). }
\end{figure}

In the main text, Eq.~(\ref{eq1-13}) describes the evolution of
the system under an effective non-Hermitian Hamiltonian, which
represents a partial scenario of open quantum system dynamics (in
the absence of quantum jumps). Generally, the complete dynamics of
the system is governed by the following Markovian master equation
(with $\hbar = 1$)
\begin{eqnarray}\label{eqc-1}
\dot{\rho(t)}&=&-i[H_{0}(t)+
\lambda\sigma_{x},\rho(t)]+{\cal{L}}[\rho(t)],\cr
{\cal{L}}[\rho(t)]&=&\gamma(t)[\sigma_{-}\rho(t)\sigma_{+}-\frac{1}{2}\{\sigma_{+}\sigma_{-},\rho(t)\},
\end{eqnarray}
with
$\gamma_{+}(t)=-\gamma_{-}(t)=\gamma(t)=\Gamma_{0}\sin^2(\frac{\omega{t}}{2})$.
In this case, the quantum state of the system is generally a mixed
state, which differs from that described in Eq.~(\ref{eq1-13}).
The population of the right eigenstate is accordingly modified as
\begin{eqnarray}\label{eqc-2}
P_{+}(t,\lambda)=\langle\widehat{\phi_{+}}(t,\lambda)|\rho(t)|\phi_{+}(t,\lambda)\rangle.
\end{eqnarray}
The above master equation is not easily solvable analytically;
therefore, we conduct a detailed numerical analysis, as shown in
Fig.~\ref{fig-Appendix_C}. The numerical results indicate that the
quantum sensor still exhibits superior performance compared to its
Hermitian counterpart ($\Gamma_{0}=0$), achieving high
susceptibility $\chi$ over a broad range of parameters.
Furthermore, variations in the dissipation coefficient can also
alter the system's sensitivity to some extent.

\end{document}